# ON RESONANT SCATTERING STATES IN GRAPHENE CIRCULAR QUANTUM DOTS

H.V. Grushevskaya[1], G.G. Krylov[1*]

[1]Physics Faculty, Belarusian State University, 4 Nezaleznasti ave., Minsk, Belarus

*e-mail: krylov@bsu.by

**Abstract.** Due to effect of Klein tunneling two-dimensional graphene quantum dots do not possess genuine bound states but quasi-bound (resonant tunneling) states only. We discuss in detail the attempt to describe these states within the framework of the Dirac pseudo-fermion model for circular dot. We demonstrate explicitly that introduced earlier the so called "resonance condition" corresponds to inconsistent system of linear equations obtained from matching conditions on the boundary of the quantum dot when one try to use it for complex energy values and to the case of total reflection for the energies coincided with the potential well top.

## 1. Introduction

Since more than twenty years of its experimental observation, graphene still attracts a lot of attention and attempts to use it as a base for super-high speed electronic devices. The masslessness of graphene carriers in the most popular graphene model, the so called pseudo-Dirac fermion model, from one side seems to be very attractive due to high value of the Fermi velocity (of about $10^7$ m/s) for the material but failed to artificially form a predefined gap in a band structure needed for transistor-like behavior.

Graphene quantum dots are considered as a variant to find a way of electric current operation. Experimentally, they can be formed with few different techniques (see, e.g. [1-3] and references therein). Theoretical description is based usually on the simplest analytically treatable case of circular symmetric quantum dot with radial step potential [4,5] (see also another approaches in [3,6] and references therein).

Specific feature of quantum problems for the 2D massless Dirac equation with the circular symmetric finite height potential barrier is the absence of bound eigenstates. This is stipulated by the fact that solutions of the Dirac radial equation in a region with a flat potential are linear combination of the Bessel functions which asymptotically look like sin and cos and therefore the eigenstates are non-normalizable and correspond to unbound particles.

In the cited references [4] it has been stated the existence of the quasi-bound eigenstates for this problem, complex energies imaginary part of which corresponds to a level's decay time. The eigenstates were chosen from some sort of a "spectral condition" arisen in consideration. In [4] and [7,8] the deduction of this condition were slightly different, in the first one it has been derived for admissible system eigenstates, in the second paper the condition was deduced for the scattering problem. In subsequent publications there were made a lot of experimental work with results interpretation based on these theoretical predictions [7,9] as well as further theoretical analysis of more complicated systems such as bilayer graphene [6], systems in an electromagnetic field [10-12].

Nevertheless, as we prove with all detail and this will be a main goal of the paper, in both cases the statement on existence of such type of quasi-bound states is erroneous. It appear due to neglecting one of two independent solutions of the corresponding radial Dirac equation when using improper physically ground assumptions.

## 2. Model

We use 2D massless Dirac fermion model of graphene [12], in tight binding approximation and near the Dirac point excitations its Hamiltonian operator reads $\hat{H}_0 = \gamma \vec{\sigma} \cdot \vec{k}$ where $\gamma$ is the band parameter linearly related with the Fermi velocity, $\vec{\sigma} = \{\sigma_x, \sigma_y\}$ is a 2D vector of two Pauli matrixes, $\vec{k}$ is the quasi-momentum.

Then, a graphene quantum dot (GQD) can be considered as graphene in some confining potential $V(\vec{r})$. The Hamiltonian reads $\hat{H} = \hat{H}_0 + V(r)$ with a scalar potential incorporated as a diagonal matrix. In the matrix form we have [5]

$$\hat{H} = \begin{pmatrix} V(\vec{r}) & \hat{p}_- \\ \hat{p}_+ & V(\vec{r}) \end{pmatrix} \qquad (1)$$

with the operators $\hat{p}_\pm$ given by $\hat{p}_- = -i\frac{\gamma}{\hbar}\frac{\partial}{\partial x} - \frac{\partial}{\partial y}$, $\hat{p}_+ = -i\frac{\gamma}{\hbar}\frac{\partial}{\partial x} + \frac{\partial}{\partial y}$.

As a model GQD we consider a circular quantum dot with a radial step potential, region $D$ is a circle with a radius $R$,

$$V(\vec{r}) = \begin{cases} 0, \vec{r} \notin D \\ V_0, \vec{r} \in D \end{cases} \qquad (2)$$

Due to the system's symmetry, for the eigenproblem $\hat{H}\psi = E\psi$, and the spinor function $\psi$ with components $\psi = (A, B)$, the separation of variables can be achieved in the polar coordinates $(\rho, \phi)$ introduced via ordinary relations $x = \rho\cos\phi, y = \rho\sin\phi$.

Then, with dimensionless variables $\rho \to \rho/R, \epsilon = RE/\gamma$ we have

$$(V(\rho) - \epsilon)A(\rho, \phi) = -ie^{-i\phi}\left(\frac{\partial B(\rho, \phi)}{\partial \rho} - \frac{i}{\rho}\frac{\partial B(\rho, \phi)}{\partial \phi}\right) \qquad (3)$$

$$(V(\rho) - \epsilon)B(\rho, \phi) = -ie^{i\phi}\left(\frac{\partial A(\rho, \phi)}{\partial \rho} + \frac{i}{\rho}\frac{\partial A(\rho, \phi)}{\partial \phi}\right) \qquad (4)$$

Substitution

$$\begin{pmatrix} A \\ B \end{pmatrix} = \begin{pmatrix} e^{im\phi} a(\rho) \\ ie^{i(m+1)\phi} b(\rho) \end{pmatrix}, \tag{5}$$

designation $\xi = \epsilon - V(\rho)$ and account of the fact that $\xi$ is different but a constant in both two regions $\rho < 1$ (inner region of the quantum dot) and $\rho > 1$ (outer region), leads to the following system of the radial equations

$$\xi a(\rho) = -b'(\rho) - \frac{(m+1)b(\rho)}{\rho} \tag{6}$$

$$\xi b(\rho) = a'(\rho) - \frac{ma(\rho)}{\rho} \tag{7}$$

Since the potential $V$ is step-like flat, expressing $b(\rho)$ from the second equation (7) of the radial-system and substituting it into the first one, we get the equation for $a(\rho)$ in the form

$$a(\rho)\left(m^2 - \xi^2 \rho^2\right) - \rho\left(\rho a''(\rho) + a'(\rho)\right) = 0 \tag{8}$$

with the general solution given by the superposition of two Bessel functions [13]

$$a(\rho) = c_1 J_m(\rho|\xi|) + c_2 Y_m(\rho|\xi|). \tag{9}$$

The last can be substituted into the equation (7) and one gets for $b(\rho) = c_1(-J_{m+1}(\rho|\xi|)) - c_2 Y_{m+1}(\rho|\xi|)$.

$$b(\rho) = c_1(-J_{m+1}(\rho|\xi|)) - c_2 Y_{m+1}(\rho|\xi|). \tag{10}$$

The boundary condition at $\rho = 0$ for the radial system demands the finiteness of the solution, so that one has to omit the second term in (10) as the function $Y_m$ is singular at zero. Outside the quantum dot ($\rho > 1$) one has to use both functions in the solution. The only requirement left is the continuity of eigenfunctions at the boundary of the quantum dot at $\rho = 1$.

Important point of the problem is that eigenfunctions can not belong to the bound state because independent of the values of the potential in both regions the asymptotically at infinity ($\rho \to \infty$) all Bessel functions trend to plane waves and therefore are non-normalizable.

Let us designate the coefficients $c$ in the solution by additional upper indexes $i$ and $o$ for inside and outside region of the quantum dot respectively. Matching solutions for both spinor components at $\rho = 1$ and choosing the normalization constant in the interior region with $c_1^i = 1$ (due to linearity of the equation one can choose arbitrary normalization) we get the following linear system for determination of the coefficients

$$J_m(|V_0 - \epsilon|) = c_1^o J_m(|\epsilon|) + c_2^o Y_m(|\epsilon|), \tag{11}$$

$$-J_{m+1}(|V_0 - \epsilon|) = c_1^o(-J_{m+1}(|\epsilon|)) - c_2^o Y_{m+1}(|\epsilon|) \qquad (12)$$

It can be shown that the determinant of the matrix of this linear system is given by the formula

$$J_{m+1}(|\epsilon|)Y_m(|\epsilon|) - J_m(|\epsilon|)Y_{m+1}(|\epsilon|), \qquad (13)$$

it does not depend on $V_0$ and is non-zero for all real values of $m$, and $\epsilon$ (see e.g., the energy dependence for $m=0$ in Fig.~1). It means that we get real valued solution of the eigenproblem which is continuous in space. As an example both spinor components of the solution are shown in Fig.~2 for $V_0 = 3, m = 0, \epsilon = 0.09$.

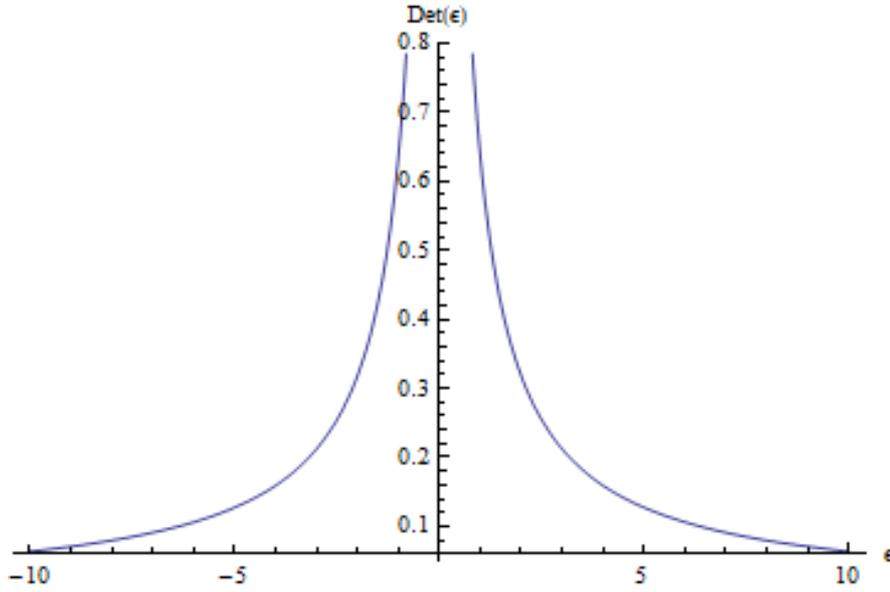

Figure 1. The dependence of the matching conditions linear system's determinant upon the energy for $m=0$

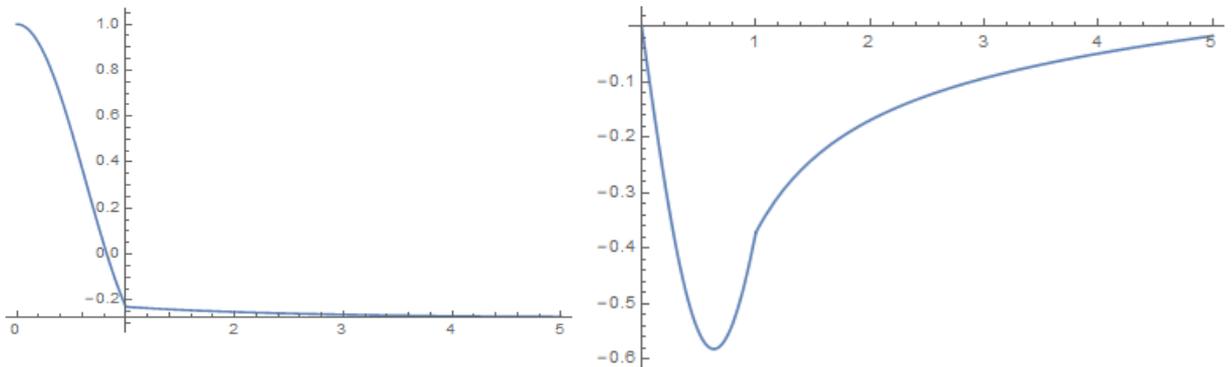

Figure 2. The upper and lower components of the spinor eigenfunction for the case $m = 0$, $\epsilon = 0.09$, $V_0 = 3$

Of course, instead of the Bessel functions $J_m, Y_m$ one can use the Hankel functions of the first and second kind $H_m^{(k)}(z) = J_m(z) \pm \imath Y_m(z), k = 1, 2$ for the outer region as it has been done in [5] resulting in complex coefficients.

But still all eigen-functions are functions of real variable and belong to continuous spectrum. In [5] authors put an additional requirement that the eigenfunction should asymptotically behaves as $\exp\{\imath|\xi|r\}$. From the physical point of view it means that one fixes the phase of the exponential function, because a linear combination of the Bessel (or Hankel) functions asymptotically can be represented as $\text{Re}(a\exp\{I kr + \phi\})$. This additional restriction leads to a specific condition which was stated in [5] as a "spectral condition"

$$J_{m+1}(|\epsilon - V_0|)H_m^{(1)}(|\epsilon|) - J_m(|\epsilon - V_0|)H_{m+1}^{(1)}(|\epsilon|) = 0 \tag{14}$$

This equation has no real solutions for $\epsilon$ except of those for $m > 1, \epsilon = 0$ but has complex ones which were interpreted in [5] as quasi-bound states with a finite lifetime.

As we intend to demonstrate the condition 14) is meaningless because corresponds to inconsistent system of linear equations followed from the matching conditions at the quantum dot boundary $\rho = 1$. With this in mind we follow the "scattering scheme" of [7,8] to obtain the matching condition system.

For radial scattering problem inside the GQD the solution consists of the transient wave and still is described by the Bessel $J$ function with some amplitude $tJ_m(|\xi|r)$ (e.g., for upper component, letter $t$ is used for transmission coefficient). Solution outside the GQD is a superposition of two Hankel functions describing incident and scattered waves that is $H_m^{(1)}(\epsilon) + rH_m^{(2)}(\epsilon)$ again for upper component (with "r" letter to designate the reflection coefficient). Then the matching system reads

$$tJ_m(|V0 - \epsilon|) - rH_m^{(1)}(|\epsilon|) - H_m^{(2)}(|\epsilon|) = 0 \tag{15}$$

$$-tJ_{m+1}(|V0 - \epsilon|) + rH_{m+1}^{(1)}(|\epsilon|) + H_{m+1}^{(2)}(|\epsilon|) = 0 \tag{16}$$

Precisely, the determinant of this system is just the l.h.s of the condition 14). So, our goal is to construct the system in the vicinity of the vanishing determinant. First we do it for the energy values $\epsilon = V_0$ corresponding to the top of the potential well. The determinant of the matrix in this case vanishes due to the properties of Bessel $J$ functions at zero argument. But we apply this energy value directly to the system to get

$$t \times 0 - rH_m^{(1)}(|V_0|) - H_m^{(2)}(|V_0|) = 0 \tag{17}$$

$$-t \times 0 + rH_{m+1}^{(1)}(|V_0|) + H_{m+1}^{(2)}(|V_0|) = 0 \tag{18}$$

The last means that the reflection coefficient has to be expressed as a ratio ($r = H_m^{(1)}(|V_0|) / H_m^{(2)}(|V_0|)$) of Hankel functions but with the orders differ on ``one'' for the first and the second equations. Taking into account the fact that the Hankel functions of the first and second kind at a given value of the argument are the conjugated complex numbers, the reflection coefficient as their ratio turns out to be unimodal. But the equality of it for both equation can be only approximately satisfied in an asymptotic regime and for small and intermediate values of the confining potential. As an example we represent in Fig. 3 the value

$\arg\left(\dfrac{H^{(1)}_{20}(x)}{H^{(2)}_{20}(x)}\right) / \arg\left(\dfrac{H^{(1)}_{21}(x)}{H^{(2)}_{21}(x)}\right)$ as a function of the potential height $V_0$ to confirm our statement.

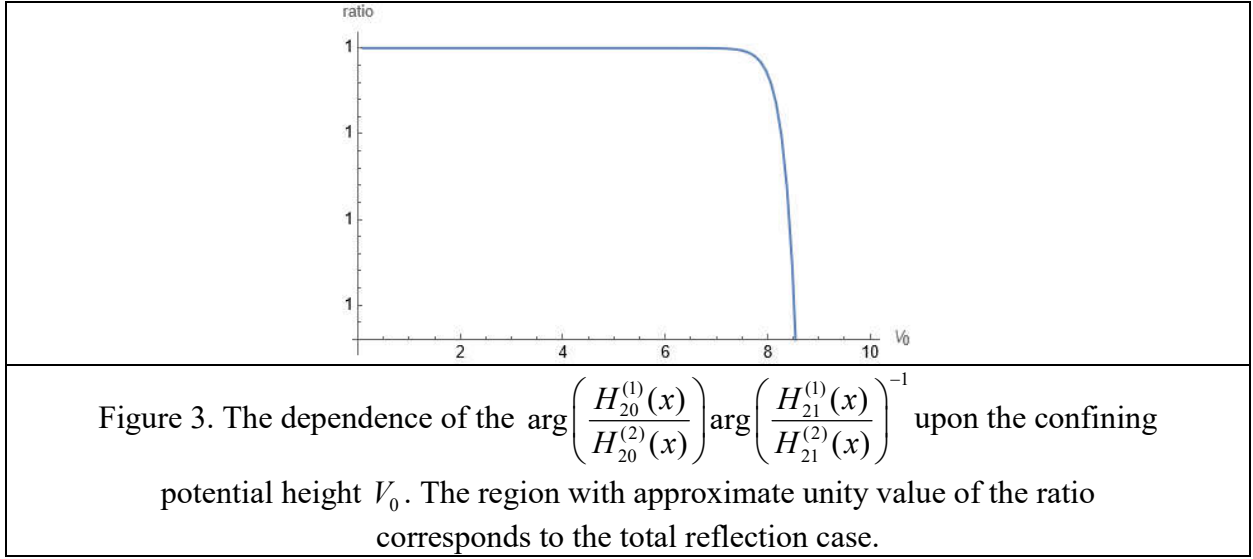

Figure 3. The dependence of the $\arg\left(\dfrac{H^{(1)}_{20}(x)}{H^{(2)}_{20}(x)}\right)\arg\left(\dfrac{H^{(1)}_{21}(x)}{H^{(2)}_{21}(x)}\right)^{-1}$ upon the confining potential height $V_0$. The region with approximate unity value of the ratio corresponds to the total reflection case.

The last regime corresponds to the same amplitudes for the reflected wave as an incident one with only a phase shift and therefore to zero value for the transmitted wave. The last means the total wave reflection case that hardly be considered as "a bound state as it was stated in [5,6].

Now, we try to understand what happens near non-trivial complex roots of the equation (14) in the case when the energy is not coincides with the well height. We choose the values for the parameters $V_0 = 20, m = 0$ and find one of the root of eqs.(17-18) in the vicinity of the $\epsilon = 3$ point. It gives $\epsilon = 3.95744 - 1.47721 I$. After substitution of this value into the matching condition system and the normalization of both equations to make coefficients at variable $t$ equal to unity we obtain

$$1.t - (1.01973 + 0.160047\,I)r = -0.0390394 + 0.0318167\,I \qquad (19)$$
$$1.t - (1.01973 + 0.160047\,I)r = 0.0409105 - 0.0408179\,I \qquad (20)$$

As one can see that indeed the determinant of the system vanishes (as both right hand sides of the system equations coincide). This means that we really have a root of the "resonance condition" but the right hand sides are different for both equations. The last shows that the system is inconsistent and there are no solutions at all in this case. The similar situation holds for other roots also.

## 3. Conclusion

Finalizing our finding. We explicitly demonstrate that the "resonance condition" introduced in [5] to construct "quasi-bound" states for GQD with a finite life time, leads to the inconsistent system of linear equations describing matching conditions for the solution inside and outside GQD. Therefore it can not be considered as valid for any application. We also demonstrate that "exact localization of electron in the quantum dot" considered in [6] and confirmed in [5] in fact can be approximately the case of the total reflection of an electron on a graphene quantum dot at asymptotically high values of magnetic quantum number $m$ and not very high value of the confining potential (whispering gallery modes).

***Acknowledgements***. *The authors are acknowledged the partial support within the projects in "Convergence-2025" and "Energetics" State Programs of the Fundamental Researches of the Republic of Belarus.*